\documentclass[runningheads]{llncs}
\usepackage[T1]{fontenc}
\usepackage{graphicx}

\usepackage[utf8]{inputenc} %
\usepackage[T1]{fontenc}    %
\usepackage{hyperref}       %
\usepackage{url}            %
\usepackage{booktabs}       %
\usepackage{amsfonts}       %
\usepackage{nicefrac}       %
\usepackage{microtype}      %
\usepackage{xcolor}         %
\usepackage{graphicx}
\usepackage{amsmath}
\usepackage{wrapfig}
\usepackage[toc,page]{appendix}
\usepackage{amssymb}
\usepackage{bbding}
\usepackage{multirow}
\usepackage{paralist}
\usepackage{mdframed}
\usepackage{framed}
\usepackage{longtable}
\usepackage[table]{xcolor}

\newcommand{\manabe}[1]{\textcolor{black}{#1}}
\newcommand{\shibata}[1]{\textcolor{black}{#1}}

\definecolor{lightgreen}{RGB}{230,255,230}

\begin{document}
\title{
Sign-to-Speech Prosody Transfer \\
via Sign Reconstruction-based GAN
}
\author{Toranosuke Manabe\orcidID{0009-0003-5058-7735} \and
Yuto Shibata\orcidID{0009-0005-4225-3887} \and
Shinnosuke Takamichi\orcidID{0000-0003-0520-7847} \and
Yoshimitsu Aoki\orcidID{0000-0001-7361-0027}
}
\authorrunning{T. Manabe et al.}
\institute{Keio University, Yagami Campus, Yokohama Kanagawa 223-8522, Japan
\email{torapira@keio.jp},
\email{yuto19990715@gmail.com},
\email{shinnosuke\_takamichi@keio.jp},
\email{aoki@elec.keio.ac.jp}
}
\maketitle              %
\begin{figure}[ht]
\centering
\vspace{-6mm}
\centerline{\includegraphics[width=\linewidth]{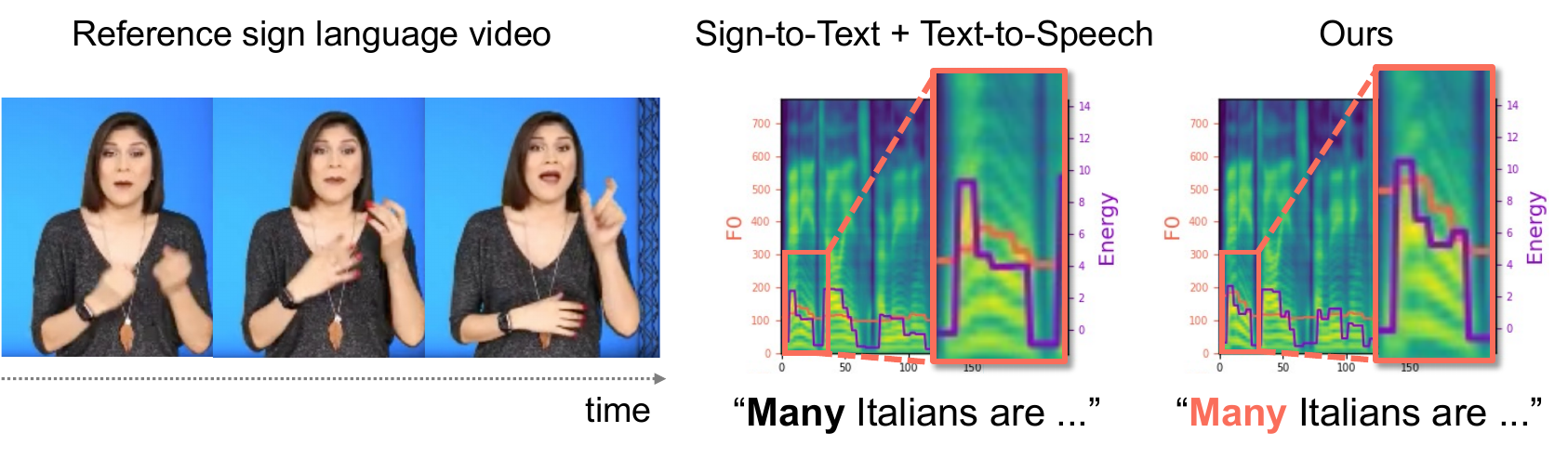}}
\vspace{-5mm}
\caption{
In the reference sign language video (left), the first phrase, ``many Italians,'' is emphasized through rapid hand movements and facial expressions. The two-stage baseline (middle) fails to reflect this prosody, whereas our approach (right) successfully captures the emphasis on ``many.''}
\vspace{-8mm}
\label{fig:pipeline}
\end{figure}

\begin{abstract}
Deep learning models have improved sign language-to-text translation and made it easier for non-signers to understand signed messages. 
When the goal is spoken communication, a naive approach is to convert signed messages into text and then synthesize speech via Text-to-Speech (TTS).
However, this two-stage pipeline inevitably treat text as a bottleneck representation, causing the loss of rich non-verbal information originally conveyed in the signing. 
To address this limitation, we propose a novel task, 
\emph{Sign-to-Speech Prosody Transfer}, 
which aims to capture the global prosodic nuances expressed
in sign language and directly integrate them into synthesized speech. 
A major challenge is that aligning sign and speech requires expert knowledge, 
making annotation extremely costly and preventing the construction of large parallel corpora. 
To overcome this, we introduce \emph{SignRecGAN}, 
a scalable training framework that leverages unimodal datasets 
without cross-modal annotations through adversarial learning 
and reconstruction losses. 
Furthermore, we propose \emph{S2PFormer}, 
a new model architecture that preserves the expressive power of existing TTS models 
while enabling the injection of sign-derived prosody into the synthesized speech.
Extensive experiments demonstrate that the proposed method can synthesize speech that faithfully reflects the emotional content of sign language, thereby opening new possibilities for more natural sign language communication. Our code will be available upon acceptance.

\end{abstract}

\section{Introduction}
\label{sec:introduction}
Sign language serves as a principal means of communication for individuals with hearing impairments.
Translating between sign language and spoken language is, 
however, a complex task that typically requires specialized knowledge.
Recently, sign-to-text translation methods based on deep learning have been developed~\cite{Camgoz_2020_CVPR,zhang2023sltunet,lin-etal-2023-gloss,Gong_2024_CVPR,NEURIPS2024_ced76a66}.
These methods substantially simplify the process of bridging sign language and spoken language.

\manabe{
However, existing studies are all limited to sign-to-text conversion.
Therefore, to convert sign language into speech using current methods, 
one must rely on a decoupled process consisting of sign-to-text and text-to-speech (TTS). 
In this pipeline, the intermediate text inevitably becomes a bottleneck, 
causing non-verbal information present in the original sign, such as tension and emphasis, to be lost.
Omitting these elements can result in speech that lacks expressiveness.
}

To fill this gap, we propose a novel task,
\emph{sign-to-speech prosody transfer}
to incorporate the global prosody embedded in sign language into speech synthesis process.
\manabe{
This task can be seen as an extension of existing 
cross-lingual prosody transfer~\cite{swiatkowski23_interspeech}.
While the existing task aims to transfer prosody 
from source spoken language to target one, 
our task transfers from source \textit{sign language}.
}

\shibata{
Here, translating between sign language and speech requires highly specialized expertise.
Consequently, constructing large-scale paired datasets is much more difficult than in other multi-modal transformation problems such as text-to-image~\cite{Rombach_2022_CVPR}, where massive paired data are readily available.
}
\manabe{ 
Moreover, both sign language and speech are time-series data, 
and creating annotations that take their temporal alignment into account is even more challenging.
}

\shibata{
To address these challenges, in this paper we propose \emph{SignRecGAN}, 
a method for converting global prosody from sign language into speech 
using separate \emph{unpaired unimodal datasets} of sign and speech,
together with \emph{S2PFormer}, a \underline{S}ign-to-\underline{P}rosody Trans\underline{former} 
designed for high-quality prosody-aware speech synthesis.  
SignRecGAN combines adversarial learning with reconstruction losses,
\emph{SignRec loss} (Sign Reconstruction loss) and \emph{ProMo loss} (Prosody-Motion alignment loss), 
which explicitly force the model to reconstruct motion information of sign language from speech, 
thereby enabling the motion expressed in signing to be reflected in natural-sounding speech. 
S2PFormer treats a large-scale pretrained TTS model as the base speech synthesis module 
and augments it with a carefully designed additional cross-attention branch, 
achieving speech synthesis that incorporates sign information 
while preserving the original high speech quality. These training strategy and model design are highly scalable, 
because \emph{SignRecGAN} leverages annotation-free unpaired unimodal datasets 
while \emph{S2PFormer} reuse a large-scale pretrained TTS model as a base speech synthesizer. 
}

We validate the effectiveness of the SignRecGAN through a qualitative evaluation, 
demonstrating that the resulting speech reflects 
a richer range of expressions compared to the conventional two-stage approach.
\manabe{CMOS} user study also confirms that our method can synthesize emotionally nuanced speech. 
In ablation studies, we show that the guidance for reconstructing sign language motion 
contributes to prosodic representation, supporting the soundness of our system design.

In summary, our main contributions are as follows:
\begin{itemize}
\item \manabe{We propose a novel task \emph{sign-to-speech prosody transfer}
to incorporate the global prosody embedded in sign language into speech synthesis process.
For this task, we introduce \emph{SignRecGAN}, 
an adversarial \shibata{and reconstruction-based learning} framework using separate \emph{unimodal datasets} of sign and speech. We also propose \emph{S2PFormer}, an augmented pretrained TTS model 
with a carefully designed additional cross-attention branch.
}
\item \shibata{For reconstruction-based learning}, we propose \emph{SignRec loss} and \emph{ProMo loss} to address the absence of directly aligned sign and speech data.
SignRec loss ensures that the synthesized speech retains the nuances of sign language by reconstructing original sign motions,
while ProMo loss aligns the distributions of sign language and synthesized speech.
These two components ensure proper integration of sign language cues into speech synthesis and enhance overall consistency.
\item Through user studies and quantitative assessments, we demonstrate that our system captures a broader range of emphatic cues compared to existing pipelines, resulting in more expressive and natural speech.
\end{itemize}

\section{Related Work}
\label{sec:related_work}
\subsection{Prosody in Sign Language}
Just as spoken language prosody conveys emphasis or emotion in a sentence, sign language also exhibits prosodic features. The prosody of sign language is expressed through  the speed and amplitude of hand movements, as well as facial expressions. Moreover, there is a correspondence between sign language motion and spoken language~\cite{Brentari2015The}. For instance, the length of a sign is similar to the duration of speech~\cite{Wilbur1999}, the peak velocity of a sign is analogous to pitch~\cite{Limousin2010}, and the movement displacement is similar to speech intensity~\cite{Wilbur1999}. However, because the correspondence in prosody between sign language and speech is extremely complex and subtle, converting prosody from sign language to speech based on predefined rules is difficult. As a result, no research currently exists that reflects sign language prosody in speech synthesis.

\subsection{Sign-to-Speech}
Sign-to-speech collectively refers to the technologies that convert sign language into speech, and in recent years, many such studies have been conducted. However, these studies mainly focus on a two-stage approach of first converting sign language to text, followed by converting text to speech. Examples include proposals of devices for sign language recognition~\cite{Zhou2020Sign-to-speech,Sharma2020Sign,Dangat2023Sign} and image recognition-based deep learning approaches that prioritize higher speeds and better accuracy~\cite{Ojha2020Sign,R2022Indian}. Consequently, there is no established method that integrates sign language prosody into speech.

\subsection{Prosody Transfer}
\manabe{
Prosody transfer refers to techniques that condition
a TTS system on a reference speech signal 
so that its prosodic characteristic, such as pitch, energy, speaking rate, and pause patterns are reflected in the synthesized output.
Depending on whether the textual content of the reference and target utterances is identical, 
different strategies have been explored. 
When the text is the same, some studies attempt fine-grained,
word-level prosody transfer by aligning prosodic patterns to the linguistic units~\cite{karlapati20_interspeech,klimkov19_interspeech}.
In contrast, when the text differs, most methods operate at the level of global prosody~\cite{pmlr-v80-skerry-ryan18a,swiatkowski23_interspeech}, 
summarizing the overall speaking style of the reference utterance into a single representation. 
The same tendency holds in cross-lingual scenarios~\cite{swiatkowski23_interspeech}, 
even when parallel translations are available: 
constructing cleanly aligned, word- or phone-level prosody labels across languages is extremely challenging,
so utterance-level global prosody transfer remains the dominant approach. 
In sign-to-speech, the primary goal is to convey the signer’s overall expressive intent,
rather than to reproduce fine-grained prosodic details.
Therefore, as a first step, we formulate our task as global prosody transfer, 
where the reference sign is summarized into an utterance-level representation that conditions speech synthesis.
}

\subsection{Dataset of Sign Language with Speech}
\label{sec:paired_dataset}
How2Sign~\cite{Duarte_2021_CVPR} is a multimodal dataset for American Sign Language (ASL) and is currently the only dataset providing a pair of ASL and English speech.
However, there are two major problems for sign-to-speech training.
First, the audio collected from YouTube is often not clean, 
and because there are numerous speakers in the dataset, 
the methods for speech modeling become limited.
Second, the dataset lacks a correspondence in prosody between the sign language and speech. 
Although the sign language was performed after viewing the original video,
six of the eleven signers in the dataset creation had hearing difficulties;
therefore, the sign language prosody does not necessarily reflect the original speech prosody.
In general, creating a paired dataset for sign language and speech requires 
careful production by annotators with specialized knowledge,
which limits both dataset scale and alignment quality.

\section{Method}
\label{sec:method}
\begin{figure}[ht]
    \centering
    \centerline{\includegraphics[width=\linewidth]{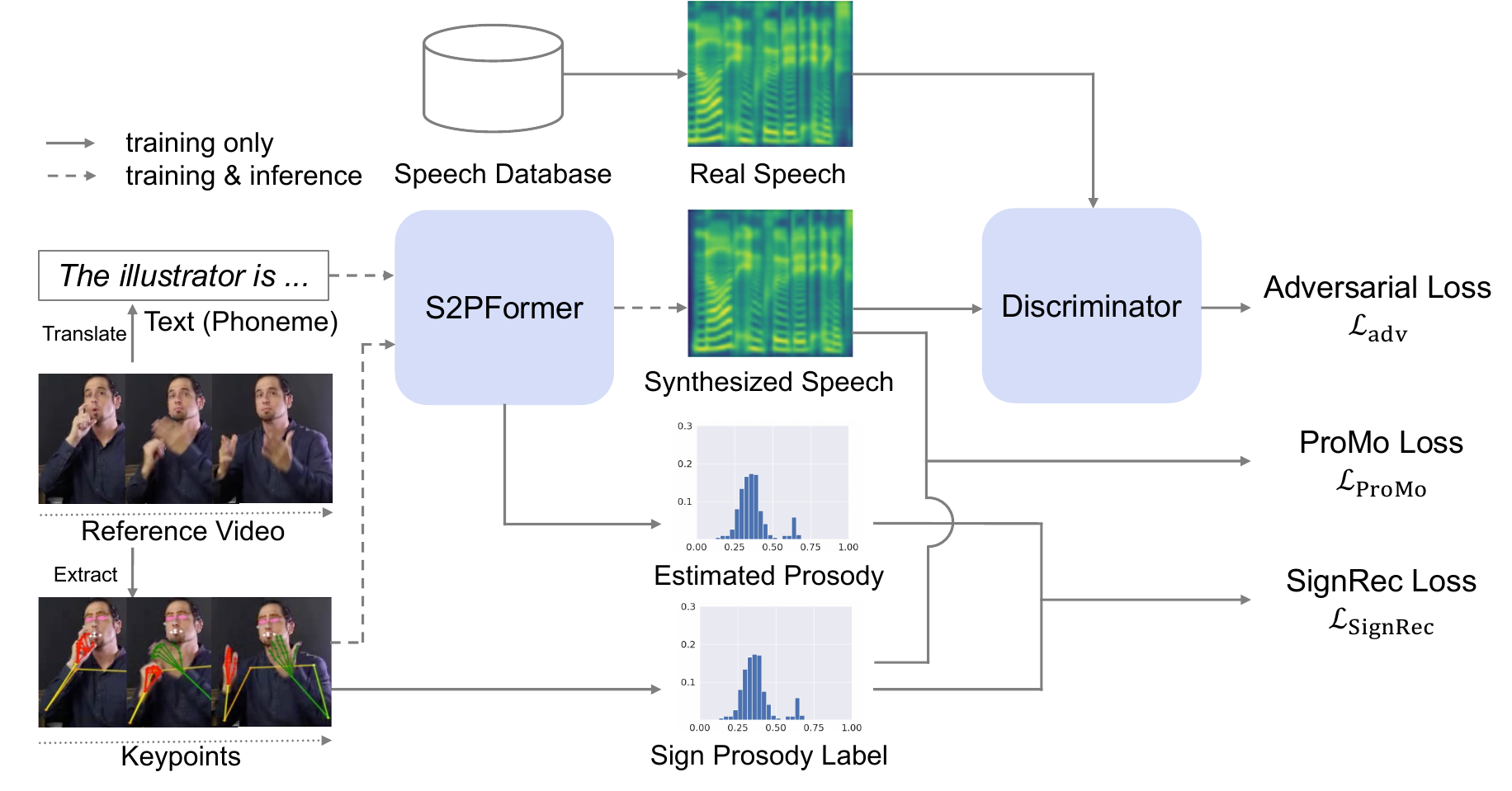}}
    \vspace{-2mm}
    \caption{The proposed learning framework of SignRecGAN.}
    \vspace{-4mm}
    \label{fig:overview}
\end{figure}

\manabe{
We propose a novel task \emph{Sign-to-Speech Prosody Transfer},
aimed at capturing the prosodic nuances conveyed
by sign language and integrating them directly
into speech synthesis.
As the first baseline for this task, 
we propose SignRecGAN, a reconstruction- and adversarial-learning framework for prosody-aware speech synthesis. 
By reconstructing motion from the synthesized speech, 
SignRecGAN encourages the synthesized speech to preserve prosodic cues originally expressed in signing.
Adversarial training further promotes naturalness while enabling these prosodic cues to be reflected in a perceptually realistic manner.
}

As shown in Fig.~\ref{fig:overview}, 
the SignRecGAN has four main components.
First, the S2PFormer (Sec~\ref{sec:ts2s}), our transformer-based model, converts a pair of text and sign language video into speech that reflects sign language prosody.
Next, we introduce SignRec loss (Sec~\ref{sec:prosody_rec}), which reconstructs sign language prosody from the predicted speech. 
Additionally, we introduce ProMo loss (Sec~\ref{sec:PGR}),
a regularization method that leverages prior knowledge about
how speech prosody corresponds to the movements in sign language.
Finally, adversarial learning ensures that the synthesized speech remains realistic (Sec~\ref{sec:training_and_inference}).
\begin{figure}[t]
    \centering
    \centerline{\includegraphics[width=\linewidth]{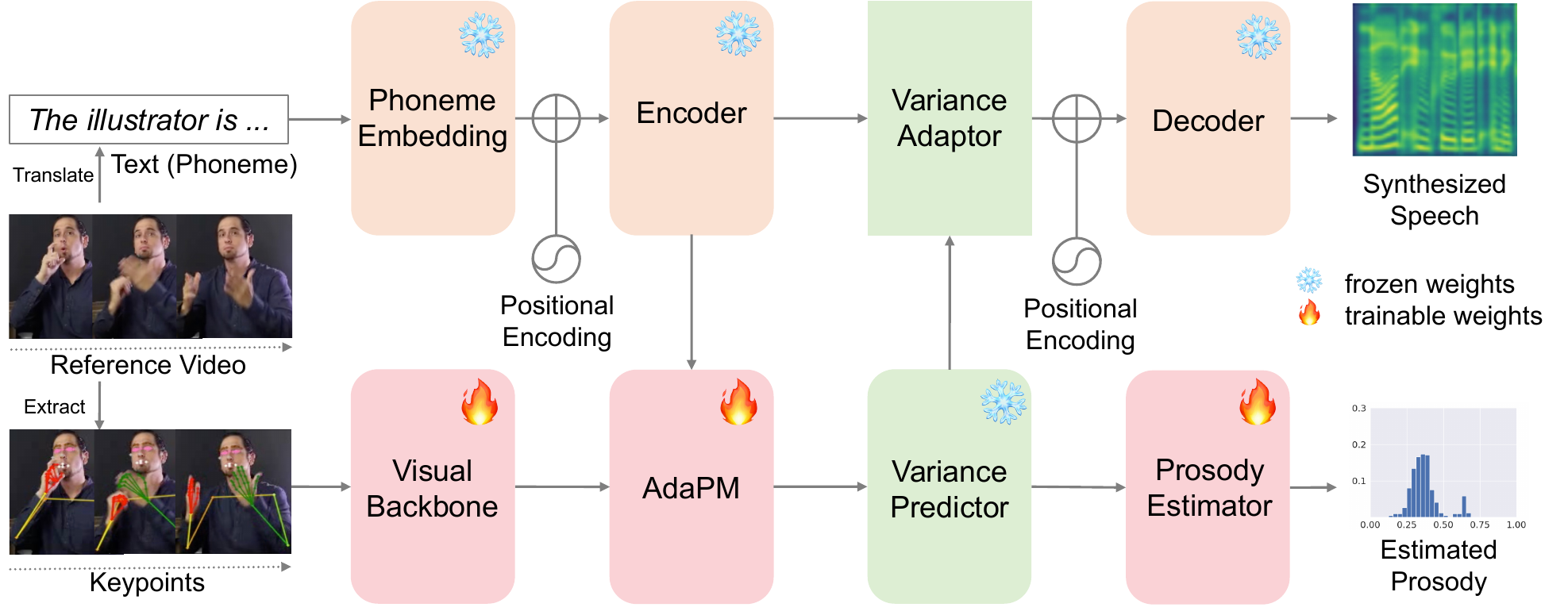}}
    \vspace{-2mm}
    \caption{The architecture of S2PFormer. S2PFormer extends FastSpeech2 by incorporating sign language information through a module called AdaPM. Specifically, the visual backbone converts sign language inputs into feature representations, which are then fed into AdaPM. Conditioned on these representations, the variance predictor estimates speech prosody parameters, which the prosody estimator uses to predict the original sign language prosody.}
    \vspace{-4mm}
    \label{fig:ts2s}
\end{figure}

\subsection{S2PFormer}
\label{sec:ts2s}
The goal of S2PFormer is to effectively extract prosodic information from sign language and integrate it into speech synthesis to synthesize speech enriched with emotion and emphasis.
This module takes as input a sign language video and its translated text sequence, 
and outputs a mel-spectrogram that reflects the prosody of sign language.
We must preserve the original linguistic content for high-quality speech synthesis. 
\manabe{
We adopt a pre-trained frozen FastSpeech2~\cite{ren2021fastspeech} backbone to maintain linguistic accuracy. Additionally,  we use a visual backbone, Adaptive Prosody Mixer (AdaPM) and a prosody estimator to incorporate sign language information.
}
This approach allows us to learn the relationship between sign language and speech prosody without compromising speech quality.

\noindent
\textbf{Preliminaries of text-to-speech synthesis.}
As illustrated in Fig. \ref{fig:ts2s}, a typical FastSpeech2-based module consists of four main components: 
encoder, variance predictor, variance adaptor, and decoder.
Specifically, the encoder first converts a phoneme sequence into phoneme latent representations. 
Then, the variance predictor estimates speech prosody information, pitch, energy and duration, from the phoneme latent representations. 
These predicted prosodic features and phoneme latent representations are passed to the variance adaptor, 
and then processed by the decoder to synthesize a mel-spectrogram.
In the variance adaptor, latent vectors corresponding to pitch and energy are 
first added to phoneme latent representations. 
Then, each element of phoneme latent representations is repeated according to predicted duration.

To incorporate sign language information into speech synthesis, 
we introduce three additional modules: 
visual backbone, Adaptive Prosody Mixer (AdaPM), and prosody estimator.
These components enable the integration of sign language information into speech prosody modeling, enhancing the prosodic naturalness of the synthesized speech.

\noindent{\textbf{Visual Backbone.}}
Sign language is a visual language composed of both motion-based and posture-based expressions.
To extract more abstract features that capture relationships among keypoints and their temporal dynamics, we adopt the keypoint feature extractor from GloFE~\cite{lin-etal-2023-gloss}.
This visual backbone combines CTR-GCN~\cite{Chen_2021_ICCV}, which models keypoint relations as a graph, and MS-TCN~\cite{Liu_2020_CVPR}, which models keypoint positions across frames at multiple temporal resolutions. This backbone transforms sign language keypoints into features.

\begin{figure}[th]
    \centering
    \vspace{-2mm}
    \centerline{\includegraphics[width=0.8\linewidth]{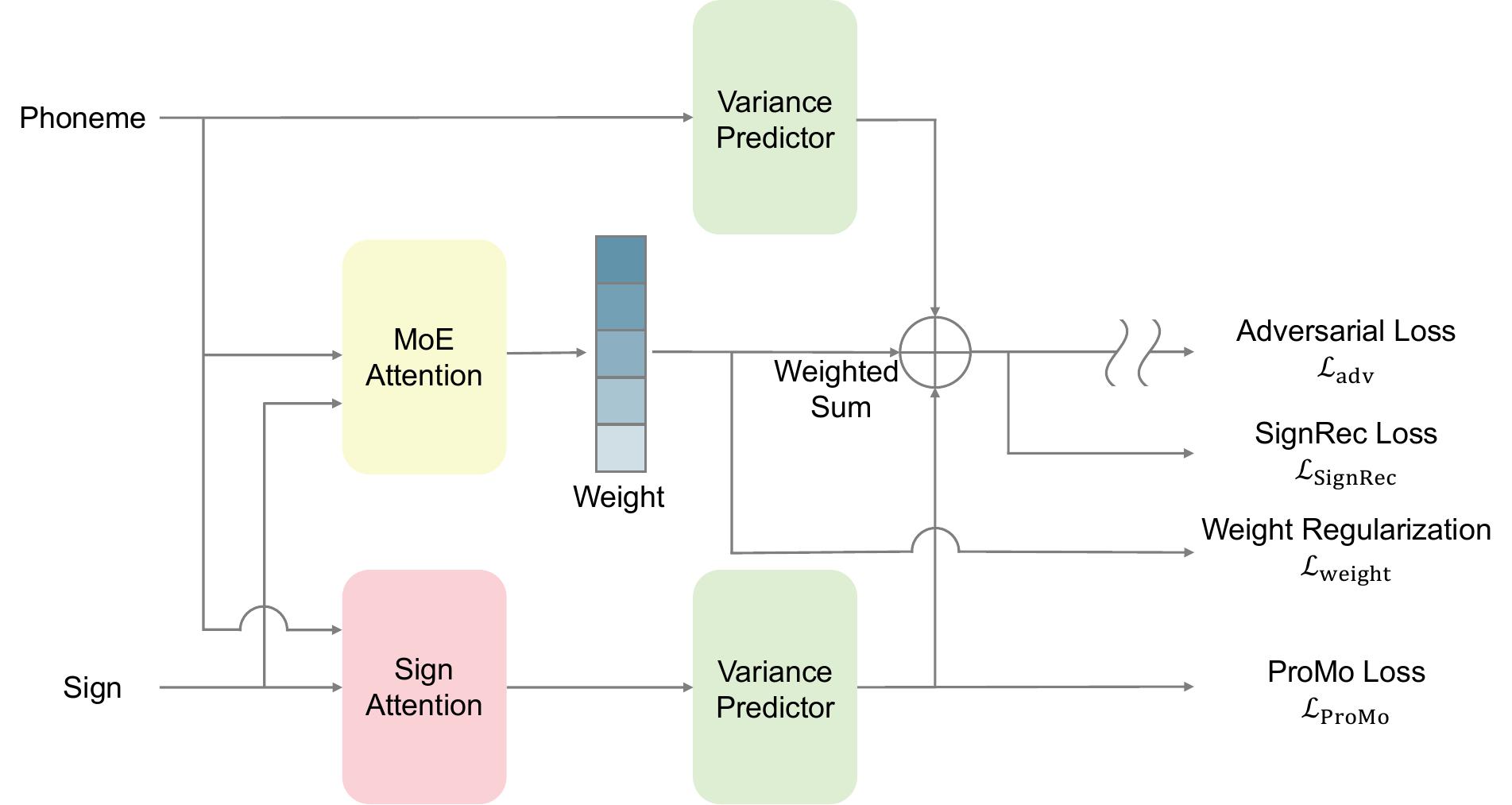}}
    \vspace{-2mm}
    \caption{Adaptive Prosody Mixer}
    \label{fig:adapm}
\end{figure}

\noindent{\textbf{Adaptive Prosody Mixer (AdaPM).}}
To reflect sign language information in speech prosody while preserving the speech reality, we introduce AdaPM inspired by Mixture of Experts (MoE)~\cite{shazeer2017outrageouslylargeneuralnetworks}. As shown in Fig.~\ref{fig:adapm}, 
this module adaptively mixes the variance predictions from the phoneme and the sign to balance \manabe{the speech naturalness and the introduced prosody from the reference sign}.
Sign attention and MoE attention are based on the transformer decoder structure~\cite{NIPS2017_3f5ee243}, of which learnable parameters are initialized to zero, 
preventing the speech prosody from becoming noisy at the early stages of training and stabilizing adversarial learning.
Moreover, to avoid the weight for sign language features being too small, we introduce a regularization term defined as follows:
\begin{equation}
    \mathcal{L}_{\mathrm{weight}} = |1 - w_{\mathrm{sign}}|
\end{equation}
where $w_{\mathrm{sign}}$ is the mean of weight for sign language features, and $\lambda_{\mathrm{weight}}$ is a hyperparameter that controls the strength of this regularization.
\begin{figure}[t]
    \centering
    \centerline{\includegraphics[width=0.8\linewidth]{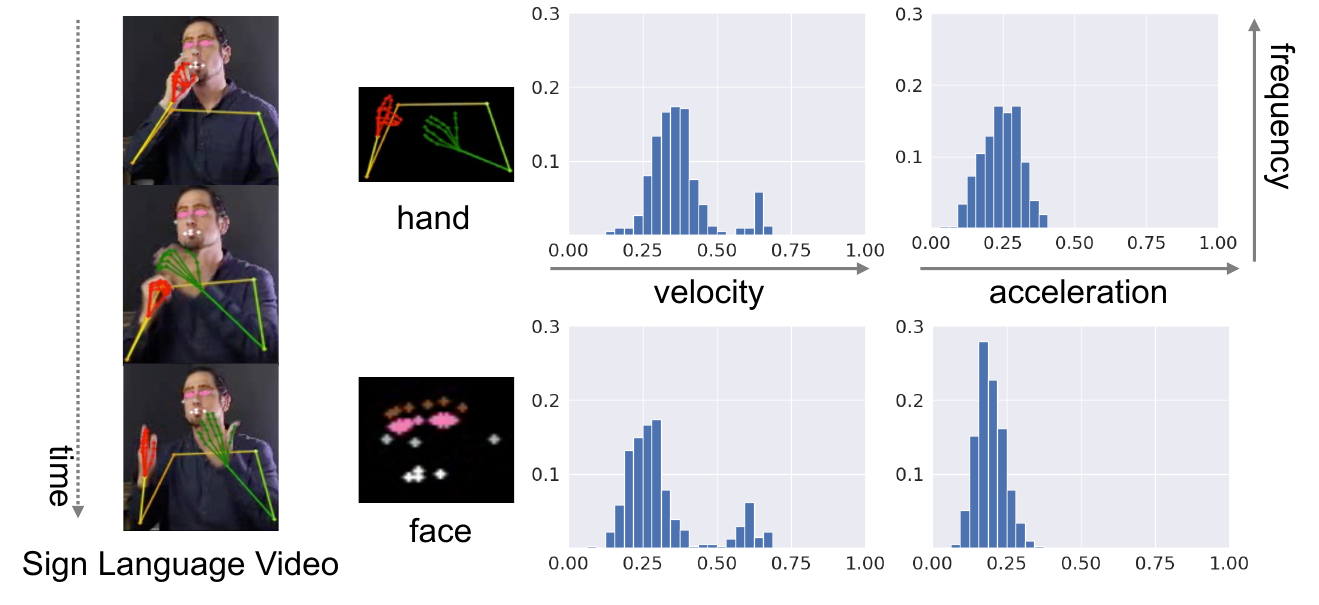}}
    \vspace{-2mm}
    \caption{An example of sign language prosody labels. The histograms represent the distribution of hand and face motion information in sign language videos.}
    \vspace{-4mm}
    \label{fig:sign_label}
\end{figure}

\subsection{SignRec Loss}
\label{sec:prosody_rec}
To ensure that the speech synthesized by S2PFormer (Sec.~\ref{sec:ts2s}) reflects sign language prosody,
we introduce the SignRec loss, which encourages the prosody estimator to reconstruct sign language prosody from the resulting speech prosody.
If sign language prosody cannot be inferred at all from the speech prosody, 
it indicates there is no correlation between the two.

\noindent
{\textbf{Sign language prosody label.}} In SignRecGAN, 
sign language prosody is defined as the distribution of motion information, 
specifically the velocity and acceleration of representative points of the hands and face in sign language videos (see Fig.~\ref{fig:sign_label}).
To create effective sign language prosody labels, 
we first define the sets of representative points for the hands and face, 
denoted as $\mathcal{H}_{\mathrm{hand}}$ and $\mathcal{H}_{\mathrm{face}}$. 
The velocity $v^i_t=p^i_{t+1}-p^i_t$ and acceleration $a^i_t=v^i_{t+1}-v^i_t$ are then computed for each keypoint $i$. 
\manabe{
To treat each motion information as a scalar value, the squared sum of the values for each representative point set is used,
that is, $V^b_t=\sum_{i\in\mathcal{H}_{b}}\|v^i_t\|^2$, 
$A^b_t=\sum_{i\in\mathcal{H}_{b}}\|a^i_t\|^2$,
$b \in \{\mathrm{hand}, \mathrm{face}\}$. 
}
Let $\mathcal{M}$ be the set of pairs, 
each consisting of a body part and its corresponding motion information. 
For each $M \in \mathcal{M}$, we denote by $M_t$ the squared magnitude of the motion corresponding to $M$ at time $t$.
Next, the squared sums of each motion information are converted into histograms:
\begin{equation}
H_{M}(k)=\sum_{t}\iota(s_k\le M_t <s_{k+1}), s_k=k/S, 0 \le k \le S
\end{equation}
where $\iota(\bullet)$ is an indicator function, and $s_k$ represents the histogram boundaries. The sign language prosody labels are then obtained by normalizing these histograms by the sequence length $T$:
\begin{equation}
P_{M}(k)=\frac{1}{T}H_{M}(k)
\end{equation}

\noindent
{\textbf{Prosody Estimator.}}
The predicted prosody features from the variance predictor, pitch $\hat{p}_{\mathrm{pitch}}$ and energy $\hat{p}_{\mathrm{energy}}$, are concatenated and used as input to estimate the sign language prosody labels $\hat{P}_{M}$. Our SignRec loss, a loss function for the prosody estimator, is defined using cross-entropy loss as follows:
\begin{equation}\label{eq:sign_rec}
\mathcal{L}_{\mathrm{SignRec}}=-\frac{1}{4}\sum_{M \in \mathcal{M}}\sum_{k}P_{M}(k)\log\hat{P}_{M}(k)
\end{equation}
By minimizing this loss, the synthesized speech prosody can reconstruct 
the speed/acceleration variations and shifts in sign language motion,
enabling a more faithful integration 
of sign language prosody into the synthesized speech.

\subsection{ProMo Loss}
\label{sec:PGR}
The learning methods described so far do not rely on paired data, 
meaning that how sign language prosody is reflected in speech prosody remains undetermined. 
To stabilize learning, we introduce cross-modal regularization terms 
based on prior knowledge of prosody in speech and sign language. 
Building on findings in the prosody literature~\cite{Wilbur1999,Limousin2010}, we exploit two correlations: the magnitude of hand movements with speech energy, and the speed of facial movements with speech pitch.
The ProMo loss regularizes prosody by minimizing the L1 distance between speaker-wise standardized speech prosody $\mu_{\mathrm{energy}}, \mu_{\mathrm{pitch}}$ and video-wise standardized sign language prosody $\mu_{v, \mathrm{hand}}, \mu_{v, \mathrm{face}}$,
which is defined as follows:
\begin{equation}\label{eq:pro_mo}
    \mathcal{L}_{\mathrm{ProMo}}=\max\left(\left|\mu_{\mathrm{energy}} - \mu_{v, \mathrm{hand}}\right| - c, 0\right) + \max\left(\left|\mu_{\mathrm{pitch}} - \mu_{v, \mathrm{face}}\right| - c, 0 \right)
\end{equation}
where $c$ is a margin to clip the loss to zero if the two modalities are close enough.

\subsection{Training and Inference}
\label{sec:training_and_inference}
In addition to the learning strategies described in Sec.~\ref{sec:prosody_rec} and Sec.~\ref{sec:PGR}, for incorporating sign language prosody into speech, adversarial learning is applied to the synthesized mel-spectrograms to enhance the quality of the synthesized speech. Specifically, the discriminator $\mathcal{D}$ is trained to distinguish whether the input mel-spectrogram is real or synthesized, while the S2PFormer generator $\mathcal{G}$ is trained to deceive the discriminator into classifying synthesized speech as real speech. The adversarial learning loss is designed using the least squares loss from LSGAN~\cite{Mao_2017_ICCV}.
The loss function for the discriminator given a sign language input $X$ and a real speech mel-spectrogram $I$ is defined as follows:
\begin{equation}
    \mathcal{L}_{\mathrm{dis}} = \frac{1}{2}\left(|1 - \mathcal{D}(\mathcal{G}(X))|^2 + |\mathcal{D}(I)|^2\right)
\end{equation}
The adversarial loss given a sign language input $X$ is defined as follows:
\begin{equation}
    \mathcal{L}_{\mathrm{adv}} = |1 - \mathcal{D}(\mathcal{G}(X))|^2
\end{equation}
In addition to adversarial generative loss, to keep synthesized speech more natural, we introduce two regularization terms for synthesized speech: (i) regularization on the intonation, (ii) regularization on speaker likeness. The regularization on the intonation penalizes the synthesized speech if the standard deviation of the pitch and energy, $\sigma_{p}^{g}$ and $\sigma_{e}^{g}$, are smaller than those of the speech synthesized by the two-stage method, $\sigma_{p}^{t}$ and $\sigma_{e}^{t}$. The regularization is defined as follows:
\begin{equation}
    \mathcal{L}_{\mathrm{IR}} = \max(\sigma_{p}^{g} - \sigma_{p}^{t}, 0) + \max(\sigma_{e}^{g} - \sigma_{e}^{t}, 0)
\end{equation}
The regularization on the range of pitch and energy penalizes the synthesized speech if the mean of the pitch and energy, $\mu_{p}^{g}$ and $\mu_{e}^{g}$, fall outside the speaker-specific interval, $\mu_{p}^{\mathrm{speaker}} \pm 3\sigma_{p}^{\mathrm{speaker}}$ and $\mu_{e}^{\mathrm{speaker}} \pm 3\sigma_{e}^{\mathrm{speaker}}$, where $\mu_{p}^{\mathrm{speaker}}, \mu_{e}^{\mathrm{speaker}}, \sigma_{p}^{\mathrm{speaker}}, \sigma_{e}^{\mathrm{speaker}}$ are the mean and standard deviation of pitch and energy. This constraint keeps the overall pitch and energy contours from shifting wholesale, thereby preserving the original speaker’s vocal identity instead of drifting toward a different-sounding voice. The regularization on speaker likeness is defined as follows:
\begin{equation}
    \mathcal{L}_{\mathrm{SL}} = \max(|\mu_{p}^{g} - \mu_{p}^{\mathrm{speaker}}| - 3\sigma_{p}^{\mathrm{speaker}}, 0) + \max(|\mu_{e}^{g} - \mu_{e}^{\mathrm{speaker}}| - 3\sigma_{e}^{\mathrm{speaker}}, 0)
\end{equation}
The generative loss and the regularization terms are summed up as follows:
\begin{equation}\label{eq:natural}
    \mathcal{L}_{\mathrm{natural}} = \mathcal{L}_{\mathrm{adv}} + \lambda_{\mathrm{IR}}\mathcal{L}_{\mathrm{IR}} + \lambda_{\mathrm{SL}}\mathcal{L}_{\mathrm{SL}} 
\end{equation}
where $\lambda_{\mathrm{IR}}$ and $\lambda_{\mathrm{SL}}$ are weight hyperparameters for each loss.

Finally, the generator's loss function is defined as follows:
\begin{equation}
    \mathcal{L} = \mathcal{L}_{\mathrm{natural}}+\lambda_{\mathrm{SignRec}}\mathcal{L}_{\mathrm{SignRec}}+\lambda_{\mathrm{ProMo}}\mathcal{L}_{\mathrm{ProMo}} + \lambda_{\mathrm{weight}} \mathcal{L}_{\mathrm{weight}}
\end{equation}
where $\lambda_{\mathrm{SignRec}}$ and $\lambda_{\mathrm{ProMo}}$ are weights for each loss.

\section{Experiments}
\label{sec:experiments}

\subsection{Experimental Setup}

\subsubsection{Dataset}
\label{subsec:datasets}
As described in Sec~\ref{sec:paired_dataset}, a high-quality dataset that pairs sign language with speech containing prosodic information does not currently exist. Therefore, we train our model using the two unimodal datasets described below.

\noindent{\bf OpenASL}~\cite{shi-etal-2022-open} is a large-scale American Sign Language (ASL) and text dataset collected from YouTube.
It contains diverse situational sign language videos It contains 98,417 translation pairs,
making it one of the largest publicly available ASL translation datasets. 
For our experiments, we subsampled 16,275 pairs for training to evaluate the feasibility of reflecting sign prosody in speech synthesis.
The validation and test sets are composed of 961 and 970 pairs, respectively.

\noindent{\bf VCTK}~\cite{VCTK} is an English speech dataset recorded in a studio by 109 speakers. 
We use this dataset as real speech data for adversarial learning.
The total duration of audio exceeds 44 hours, recorded at 48 kHz and converted to 16-bit.

\vspace{-2mm}
\subsubsection{Dataset Preprocessing}
Sign language videos were processed with MMPose~\cite{mmpose} to extract keypoints, and then we clipped the input video frame lengths to be between 30 and 512 frames. To exclude sign language samples with excessively long text, we removed those that deviated by more than 10 standard deviations from the mean.
For the audio, we used the Montreal Forced Aligner~\cite{mfa} to obtain 80-dimensional mel-spectrograms with a frame size of 1024 and hop size of 256.

\vspace{-2mm}
\subsubsection{Network Details}
Our model is based on FastSpeech2 and was pretrained on VCTK to acquire speaker information. We utilized open-source implementations of FastSpeech2\footnote{\url{https://github.com/ming024/FastSpeech2}} and the Visual Backbone\footnote{\url{https://github.com/HenryLittle/GloFE}}.
We trained our model on 3 NVIDIA RTX A5000 GPUs with a per-GPU batch size of 10.
For the optimization method, we used AdamW~\cite{loshchilov2018decoupled} for all models. For more details regarding hyper parameters and model implementation, please refer to the Appendix.

\vspace{-2mm}
\subsubsection{Evaluation Method}
In this paper, we evaluate the quality of synthesized speech using both objective and subjective assessments.
For the objective evaluation, 
we measure the expressiveness and naturalness of synthesized speech. 
For subjective evaluation, 
we conduct a user study and report comparative mean opinion scores (CMOS),
\manabe{following previous work of cross-lingual prosody transfer~\cite{swiatkowski23_interspeech}.}

\noindent{\bf Expressiveness}
To measure the richness of emotional expression in speech,
we calculate the standard deviations of predicted pitch and energy contours. 
Larger standard deviations indicate richer emotional expression~\cite{kharitonov-etal-2022-text}.
To evaluate the overall speech quality, 
we calculate the standard deviation across the test set, 
and to evaluate individual samples, 
we calculate the standard deviation for each speech sample and then take the mean of these values.

\noindent{\bf Naturalness}
To evaluate the quality of the synthesized speech and to show that our model does not degrade the quality of the speech, 
we use WER (Word Error Rate) and UTMOS\footnote{\url{https://github.com/sarulab-speech/UTMOS22}}~\cite{saeki22c_interspeech}. 
We use the pretrained automatic speech recognition model, 
Whisper large V3\footnote{\url{https://huggingface.co/openai/whisper-large-v3}}~\cite{radford2022whisper} to calculate WER. 
In particular, some text translations in OpenASL are inappropriate for speech synthesis: too long texts or too rare vocabularies. 
Therefore, we filtered out the text longer than 18 words and the text containing words of three or more capital letters.

\noindent{\bf User Study}
\manabe{
We conducted a CMOS test to evaluate how well and naturally
the synthesized speech matches the source sign language prosody. 
}
We grouped samples into three categories based on text length: Short (3-8 words), Medium (9-13 words), and Long (14-18 words). 
In addition, we categorized the 30 sign language videos with the lowest mean facial velocity as Neutral. 
Given that the two-stage approach yields sufficient prosody, SignRecGAN's primary role in these cases is to maintain baseline speech quality without introducing artifacts.
In the CMOS test, 17 participants\footnote{A participant accidentally took the two types of tests.} rated each sample on a five grade evaluation.
Please refer to the Appendix for more details on the user study.

\noindent{\bf Prominence Analysis}
While our proposed method primarily focuses on reconstructing global prosody, it is crucial to verify whether this also results in high-quality, fine-grained prosodic features. 
To this end, we conducted a prominence analysis using the Wavelet Prosody Toolkit\footnote{\url{https://github.com/asuni/wavelet_prosody_toolkit}}~\cite{DBLP:journals/corr/SuniAV15}. For a more detailed explanation of prominence analysis,
please refer to the Appendix.

\begin{table}[t]\fontsize{10pt}{10pt}\selectfont
    \centering
    \caption{Objective evaluation. 
    The best scores are highlighted in \textbf{bold}, and the second-best scores are \underline{underlined}.}
    \label{tab:objective}
    \vspace{-3mm}
    \begin{tabular}[t]{lccccccc}
    \toprule
      &&&& \multicolumn{2}{c}{Expressiveness} & \multicolumn{2}{c}{Naturalness} \\
    Method & $\mathcal{L}_{\mathrm{natural}}$ & $\mathcal{L}_{\mathrm{SignRec}}$ & $\mathcal{L}_{\mathrm{ProMo}}$ & Pitch($\rightarrow$) & Energy($\rightarrow$) & WER($\downarrow$) & UTMOS($\uparrow$) \\
    \midrule
    Real Speech &  &  &  & 28.9 & 25.4 & 0.04 & 4.07 \\
    \midrule
    two-stage &  &  &  & \underline{17.6} & 20.1 & \underline{0.293} & \textbf{3.52} \\
    \midrule
    \multirow{4}{*}{Ours} & \Checkmark &  &  & 48.4 & 44.6 & \textbf{0.212} & 3.29 \\
     & \Checkmark & \Checkmark &  & 44.2 & 39.7 & 0.448 & 2.95 \\
     & \Checkmark &  & \Checkmark   & 45.3 & \underline{28.6} & 0.623 & 3.03 \\
     & \Checkmark & \Checkmark & \Checkmark                  & \textbf{33.7} & \textbf{22.5} & 0.325 & \underline{3.44}\\
    \bottomrule
    \end{tabular}
    \vspace{-4mm}
\end{table}

\subsection{Objective Evaluation}
Table~\ref{tab:objective} shows the result of comparing the expressiveness of the synthesized speech and that of real speech. 
As can be seen, SignRecGAN exhibits higher expressiveness values for pitch and energy than the two-stage method, 
and these values are closer to those of real speech. 
This suggests that intonation and emphasis similar to real speech can be achieved more effectively than with the two-stage method, 
although SignRecGAN slightly degrades the naturalness of the synthesized speech compared to the two-stage method.

\subsection{Ablation Study}
To investigate the effectiveness of each component in SignRecGAN, we conducted an ablation study.
GAN, SignRec, and ProMo indicate the S2PFormer with the losses defined in Eqs.~\ref{eq:natural}, \ref{eq:sign_rec}, \ref{eq:pro_mo} respectively.
The result in Table~\ref{tab:objective} show that every component alone does not contribute to improvement of expressiveness, and the combination of all three components is necessary to achieve the best performance. 
In particular, our ablation shows that 
if either the SignRec loss or the ProMo loss is removed, 
naturalness degrades even with the adversarial loss, 
indicating that all three components are necessary to maintain natural-sounding speech.

\begin{table}[t]
    \centering
    \caption{The CMOS of the user study with 95\% confidence intervals. The scores with statistically significant improvements are highlighted in \colorbox{lightgreen}{green}.}
    \vspace{-2mm}
    \begin{tabular}{l c c c c c}
        \toprule
         & Short & Medium & Long & Neutral & Total \\
        \midrule
        Prosody & $3.12 (\pm 0.22)$ & \cellcolor{lightgreen}$3.19 (\pm 0.16)$ & \cellcolor{lightgreen} $3.19 (\pm 0.16)$ & $3.00 (\pm 0.18)$ & \cellcolor{lightgreen} $3.17 (\pm 0.11)$ \\
        Naturalness & $2.93 (\pm 0.17)$ & $2.97 (\pm 0.17)$ & $2.87 (\pm 0.15)$ & $2.96 (\pm 0.15)$ & $2.92 (\pm 0.11)$ \\
        \bottomrule
    \end{tabular}
    \label{tab:user_study}
    \vspace{-2mm}
\end{table}

\subsection{User Study}
Table~\ref{tab:user_study} shows that SignRecGAN produced results more consistent with sign language prosody than the two-stage method. In particular, the results for the Medium and Long categories show statistically significant differences, whereas the two methods do not differ significantly in naturalness.
Taken together with the quantitative evaluation results, these results indicate a slight degradation in naturalness.
Also, the scores for Neutral category did not show a statistically significant difference, indicating that SignRecGAN does not degrade the quality of synthesized speech.

\begin{figure}[t] %
    \centering
    \centerline{\includegraphics[width=\linewidth]{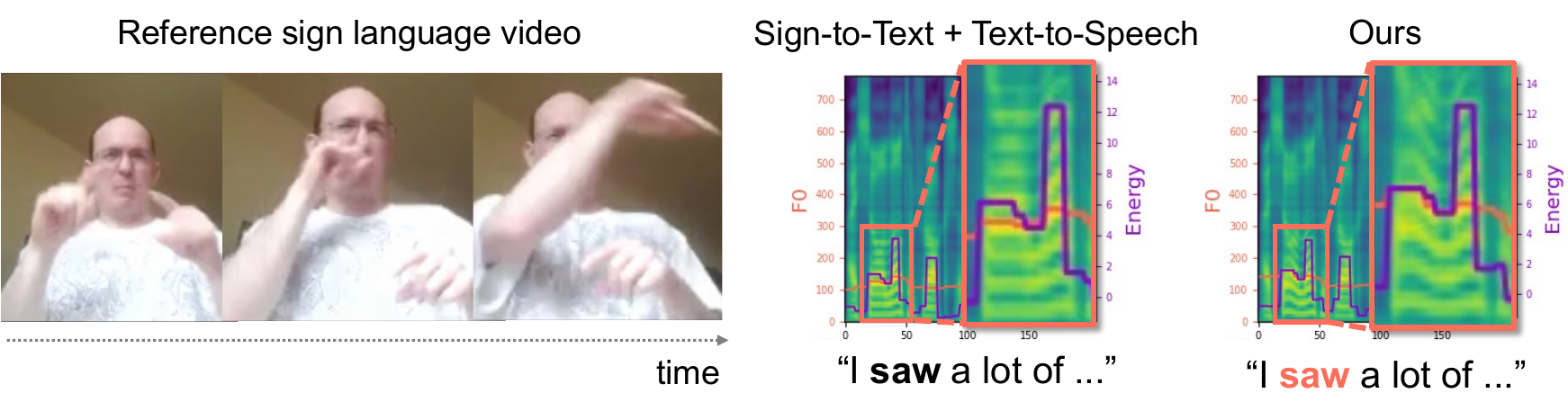}}
    \vspace{-4mm}
    \caption{An example of input sign language video (left) and synthesized speech (right).}
    \vspace{-2mm}
    \label{fig:mel_ex1}
\end{figure}

\subsection{Qualitative Evaluation}
\label{sec:qualitative}
Fig.~\ref{fig:mel_ex1} presents an example of the synthesized speech, one of the highest scored samples in the user study. In the reference sign language video, the signer signs the phrase ``I saw'' with dynamic facial expressions and hand movements. SignRecGAN successfully represents the emphasis on the phrase, while two-stage method fails.

\begin{figure}[th]
    \centering
    \centerline{\includegraphics[width=\linewidth]{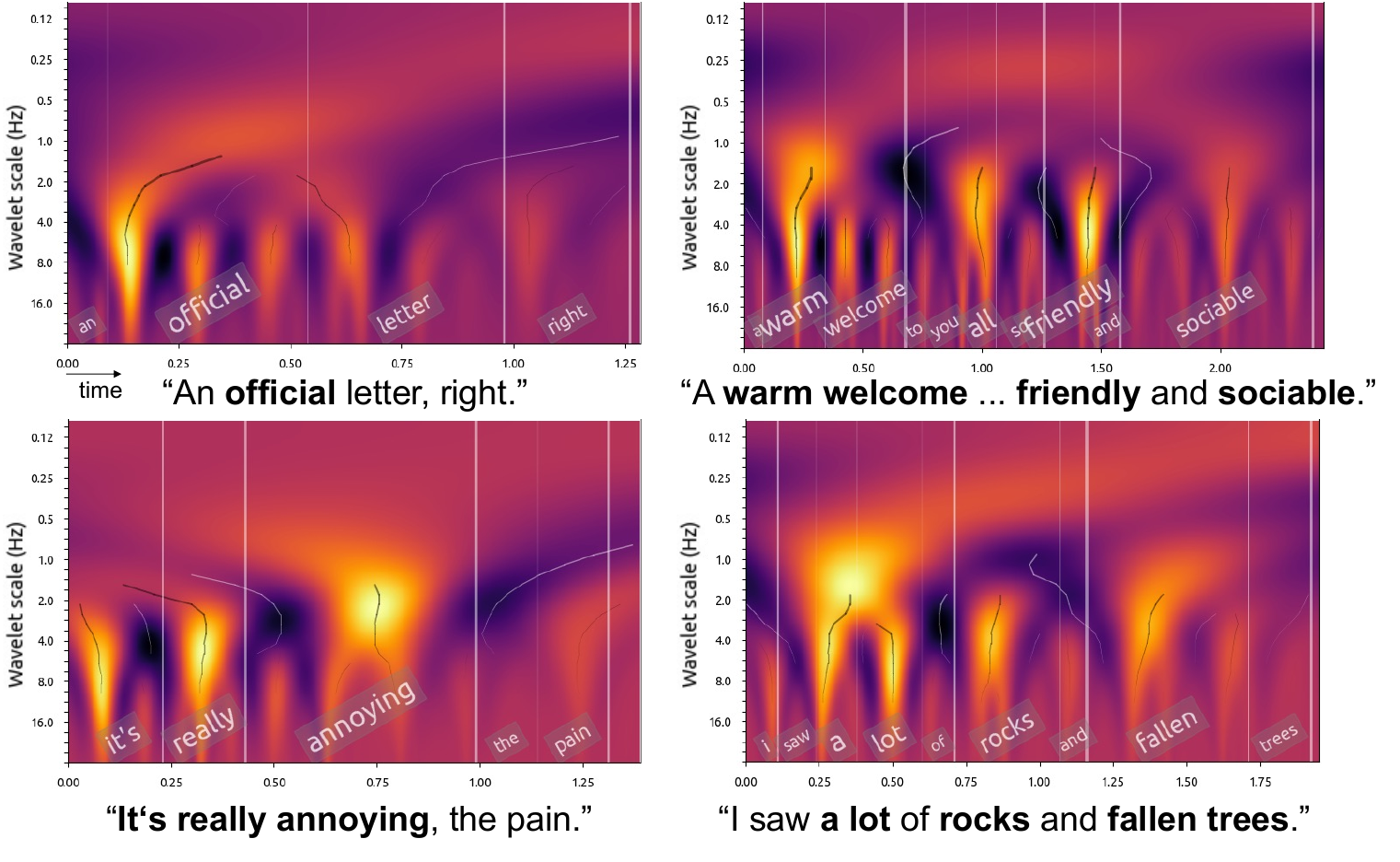}}
    \vspace{-2mm}
    \caption{Prominence analysis results. A black line on a spectrogram indicates emphasis, and the corresponding word is labeled on the spectrogram. The size of a labeled word indicates the intensity of the emphasis.}
    \label{fig:prominece}
    \vspace{-6mm}
\end{figure}

\subsection{Emergence of Emphasis on Natural Words}
\manabe{
Regarding to fine-grained prosody evaluation, Fig.~\ref{fig:prominece} shows some results of prominence analysis~\cite{DBLP:journals/corr/SuniAV15}.
The results indicate that S2PFormer can emphasize suitable words (e.g., ``\textit{annoying}'', ``\textit{a lot}'') and does not emphasize unnatural words (e.g., ``\textit{and}'').
This suggests that while the reconstruction loss facilitates the modeling of global prosody, its combination with adversarial learning enables the model to selectively emphasize contextually natural words.
}

\section{Conclusion}
\label{sec:conclusion}
In this study, we tackled a novel task \emph{Sign-to-Speech Prosody Transfer}, 
which seeks to transfer sign language prosody into synthesized speech. 
\manabe{
The major challenge for this task is the lack of datasets that align sign and speech.
Constructing such datasets requires expert knowledge, making annotation extremely costly.
To address this challenge, 
we introduce \emph{SignRecGAN}, 
a training framework utilizing unimodal datasets of speech and sign language 
through adversarial learning and reconstruction loss.
}
Through quantitative evaluation and user studies, 
we demonstrated that the proposed method can produce speech 
that better reflects the sign language prosody.

{
    \small
    \bibliographystyle{splncs04}
    \bibliography{main}

@String(CVPR= {IEEE Conf. Comput. Vis. Pattern Recog.})

@String(ICCV= {Int. Conf. Comput. Vis.})

@String(CVPR  = {CVPR})

@String(ICCV  = {ICCV})

@InProceedings{Camgoz_2020_CVPR,
    author = {Camgoz, Necati Cihan and Koller, Oscar and Hadfield, Simon and Bowden, Richard},
    title = {Sign Language Transformers: Joint End-to-End Sign Language Recognition and Translation},
    booktitle = {CVPR},
    month = {June},
    year = {2020}
}

@inproceedings{
    zhang2023sltunet,
    title={{SLTUNET}: A Simple Unified Model for Sign Language Translation},
    author={Biao Zhang and Mathias M{\"u}ller and Rico Sennrich},
    booktitle={The Eleventh International Conference on Learning Representations },
    year={2023},
    url={https://openreview.net/forum?id=EBS4C77p_5S}
}

@InProceedings{Gong_2024_CVPR,
    author    = {Gong, Jia and Foo, Lin Geng and He, Yixuan and Rahmani, Hossein and Liu, Jun},
    title     = {LLMs are Good Sign Language Translators},
    booktitle = {CVPR},
    month     = {June},
    year      = {2024},
    pages     = {18362-18372}
}

@inproceedings{NEURIPS2024_ced76a66,
    author = {Zhang, Biao and Tanzer, Garrett and Firat, Orhan},
    booktitle = {Advances in Neural Information Processing Systems},
    editor = {A. Globerson and L. Mackey and D. Belgrave and A. Fan and U. Paquet and J. Tomczak and C. Zhang},
    pages = {114018--114047},
    publisher = {Curran Associates, Inc.},
    title = {Scaling Sign Language Translation},
    url = {https://proceedings.neurips.cc/paper_files/paper/2024/file/ced76a666704e381c3039871ffe558ee-Paper-Conference.pdf},
    volume = {37},
    year = {2024}
}

@article{Brentari2015The,
  title={The acquisition of prosody in American Sign Language},
  author={D. Brentari and Joshua Falk and George Wolford},
  journal={Language},
  year={2015},
  volume={91},
  pages={e144 - e168},
  doi={10.1353/LAN.2015.0042}
}

@article{Wilbur1999,
  author = {Wilbur, Ronnie},
  year = {1999},
  month = {04},
  pages = {229-50},
  title = {Stress in- ASL: Empirical evidence and linguistic issues},
  volume = {42 ( Pt 2-3)},
  journal = {Language and speech}
}

@article{Limousin2010,
  author = {Limousin, Fanny and Blondel, Marion},
  year = {2010},
  month = {01},
  pages = {},
  title = {Prosodie et acquisition de la langue des signes française},
  journal = {Language, Interaction and Acquisition}
}

@inproceedings{
  ren2021fastspeech,
  title={FastSpeech 2: Fast and High-Quality End-to-End Text to Speech},
  author={Yi Ren and Chenxu Hu and Xu Tan and Tao Qin and Sheng Zhao and Zhou Zhao and Tie-Yan Liu},
  booktitle={International Conference on Learning Representations},
  year={2021},
  url={https://openreview.net/forum?id=piLPYqxtWuA}
}

@inproceedings{swiatkowski23_interspeech,
  title     = {Cross-lingual Prosody Transfer for Expressive Machine Dubbing},
  author    = {Jakub Swiatkowski and Duo Wang and Mikolaj Babianski and Patrick {Lumban Tobing} and Ravichander Vipperla and Vincent Pollet},
  year      = {2023},
  booktitle = {Interspeech 2023},
  pages     = {4838--4842},
  doi       = {10.21437/Interspeech.2023-437},
  issn      = {2958-1796},
}

@article{Zhou2020Sign-to-speech,title={Sign-to-speech translation using machine-learning-assisted stretchable sensor arrays},author={Zhihao Zhou and Kyle Chen and Xiaoshi Li and Songlin Zhang and Yufen Wu and Yihao Zhou and Keyu Meng and Chenchen Sun and Qiang He and Wenjing Fan and Endong Fan and Zhiwei Lin and Xulong Tan and W. Deng and Jin Yang and Jun Chen},journal={Nature Electronics},year={2020},volume={3},pages={571 - 578},doi={10.1038/s41928-020-0428-6}}

@article{Sharma2020Sign,title={Sign Language to Speech Translation},author={Aishwarya Sharma and Sibaji Panda and Saurav Verma},journal={2020 11th International Conference on Computing, Communication and Networking Technologies (ICCCNT)},year={2020},pages={1-8},doi={10.1109/ICCCNT49239.2020.9225422}}

@article{Dangat2023Sign,title={Sign Language to Speech Conversion},author={Prof. M. T. Dangat},journal={International Journal for Research in Applied Science and Engineering Technology},year={2023},doi={10.22214/ijraset.2023.56174}}

@article{Ojha2020Sign,title={Sign Language to Text and Speech Translation in Real Time Using Convolutional Neural Network},author={A. Ojha and Ayush Pandey and Shubham Maurya and A. Thakur and P. Dayananda},journal={International journal of engineering research and technology},year={2020},volume={8},doi={}}

@article{R2022Indian,title={Indian Sign Language to Speech Conversion Using Convolutional Neural Network},author={S. R and Surendra R Hegde and Chinmaya K and Ankit Priyesh and A. S. Manjunath and B. Arunakumari},journal={2022 IEEE 2nd Mysore Sub Section International Conference (MysuruCon)},year={2022},pages={1-5},doi={10.1109/MysuruCon55714.2022.9972574}}

@InProceedings{Duarte_2021_CVPR,
  author    = {Duarte, Amanda and Palaskar, Shruti and Ventura, Lucas and Ghadiyaram, Deepti and DeHaan, Kenneth and Metze, Florian and Torres, Jordi and Giro-i-Nieto, Xavier},
  title     = {How2Sign: A Large-Scale Multimodal Dataset for Continuous American Sign Language},
  booktitle = {CVPR},
  month     = {June},
  year      = {2021},
  pages     = {2735-2744}
}

@inproceedings{lin-etal-2023-gloss,
    title = "Gloss-Free End-to-End Sign Language Translation",
    author = "Lin, Kezhou  and Wang, Xiaohan  and Zhu, Linchao  and Sun, Ke  and Zhang, Bang  and Yang, Yi",
    booktitle = "Proceedings of the 61st Annual Meeting of the Association for Computational Linguistics (Volume 1: Long Papers)",
    month = jul,
    year = "2023",
    address = "Toronto, Canada",
    publisher = "Association for Computational Linguistics",
    url = "https://aclanthology.org/2023.acl-long.722",
    doi = "10.18653/v1/2023.acl-long.722",
    pages = "12904--12916",
}

@InProceedings{Chen_2021_ICCV,
    author    = {Chen, Yuxin and Zhang, Ziqi and Yuan, Chunfeng and Li, Bing and Deng, Ying and Hu, Weiming},
    title     = {Channel-Wise Topology Refinement Graph Convolution for Skeleton-Based Action Recognition},
    booktitle = {Proceedings of the IEEE/CVF International Conference on Computer Vision (ICCV)},
    month     = {October},
    year      = {2021},
    pages     = {13359-13368}
}

@InProceedings{Liu_2020_CVPR,
  author = {Liu, Ziyu and Zhang, Hongwen and Chen, Zhenghao and Wang, Zhiyong and Ouyang, Wanli},
  title = {Disentangling and Unifying Graph Convolutions for Skeleton-Based Action Recognition},
  booktitle = {CVPR},
  month = {June},
  year = {2020}
}

@InProceedings{pmlr-v80-skerry-ryan18a,
  title = 	 {Towards End-to-End Prosody Transfer for Expressive Speech Synthesis with Tacotron},
  author =       {Skerry-Ryan, RJ and Battenberg, Eric and Xiao, Ying and Wang, Yuxuan and Stanton, Daisy and Shor, Joel and Weiss, Ron and Clark, Rob and Saurous, Rif A.},
  booktitle = 	 {Proceedings of the 35th International Conference on Machine Learning},
  pages = 	 {4693--4702},
  year = 	 {2018},
  editor = 	 {Dy, Jennifer and Krause, Andreas},
  volume = 	 {80},
  series = 	 {Proceedings of Machine Learning Research},
  month = 	 {10--15 Jul},
  publisher =    {PMLR},
  url = 	 {https://proceedings.mlr.press/v80/skerry-ryan18a.html}
}

@inproceedings{karlapati20_interspeech,
  title     = {CopyCat: Many-to-Many Fine-Grained Prosody Transfer for Neural Text-to-Speech},
  author    = {Sri Karlapati and Alexis Moinet and Arnaud Joly and Viacheslav Klimkov and Daniel Sáez-Trigueros and Thomas Drugman},
  year      = {2020},
  booktitle = {Interspeech 2020},
  pages     = {4387--4391},
  doi       = {10.21437/Interspeech.2020-1251},
  issn      = {2958-1796},
}

@InProceedings{Mao_2017_ICCV,
  author = {Mao, Xudong and Li, Qing and Xie, Haoran and Lau, Raymond Y.K. and Wang, Zhen and Paul Smolley, Stephen},
  title = {Least Squares Generative Adversarial Networks},
  booktitle = {Proceedings of the IEEE International Conference on Computer Vision (ICCV)},
  month = {Oct},
  year = {2017}
}

@misc{shazeer2017outrageouslylargeneuralnetworks,
      title={Outrageously Large Neural Networks: The Sparsely-Gated Mixture-of-Experts Layer}, 
      author={Noam Shazeer and Azalia Mirhoseini and Krzysztof Maziarz and Andy Davis and Quoc Le and Geoffrey Hinton and Jeff Dean},
      year={2017},
      eprint={1701.06538},
      archivePrefix={arXiv},
      primaryClass={cs.LG},
      url={https://arxiv.org/abs/1701.06538}, 
}

@inproceedings{NIPS2017_3f5ee243,
  author = {Vaswani, Ashish and Shazeer, Noam and Parmar, Niki and Uszkoreit, Jakob and Jones, Llion and Gomez, Aidan N and Kaiser, \L ukasz and Polosukhin, Illia},
  booktitle = {Advances in Neural Information Processing Systems},
  editor = {I. Guyon and U. Von Luxburg and S. Bengio and H. Wallach and R. Fergus and S. Vishwanathan and R. Garnett},
  pages = {},
  publisher = {Curran Associates, Inc.},
  title = {Attention is All you Need},
  url = {https://proceedings.neurips.cc/paper_files/paper/2017/file/3f5ee243547dee91fbd053c1c4a845aa-Paper.pdf},
  volume = {30},
  year = {2017}
}

@inproceedings{shi-etal-2022-open,
    title = "Open-Domain Sign Language Translation Learned from Online Video",
    author = "Shi, Bowen  and
      Brentari, Diane  and
      Shakhnarovich, Gregory  and
      Livescu, Karen",
    editor = "Goldberg, Yoav  and
      Kozareva, Zornitsa  and
      Zhang, Yue",
    booktitle = "Proceedings of the 2022 Conference on Empirical Methods in Natural Language Processing",
    month = dec,
    year = "2022",
    address = "Abu Dhabi, United Arab Emirates",
    publisher = "Association for Computational Linguistics",
    url = "https://aclanthology.org/2022.emnlp-main.427/",
    doi = "10.18653/v1/2022.emnlp-main.427",
    pages = "6365--6379",
}

@misc{VCTK,
  title = "CSTR VCTK Corpus: English Multi-speaker Corpus for CSTR Voice Cloning Toolkit",
  author = "Yamagishi, Junichi and Veaux, Christophe and MacDonald, Kirsten",
  year = "2017",
  doi="10.7488/ds/1994",
}

@misc{mmpose,
  title = "OpenMMLab Pose Estimation Toolbox and Benchmark",
  author = "MMPose Contributors",
  year = "2020",
  url = "https://github.com/open-mmlab/mmpose"
}

@inproceedings{mfa,
  author = {McAuliffe, Michael and Socolof, Michaela and Mihuc, Sarah and Wagner, Michael and Sonderegger, Morgan},
  year = {2017},
  month = {08},
  pages = {498-502},
  title = {Montreal Forced Aligner: Trainable Text-Speech Alignment Using Kaldi},
  doi = {10.21437/Interspeech.2017-1386}
}

@inproceedings{
loshchilov2018decoupled,
title={Decoupled Weight Decay Regularization},
author={Ilya Loshchilov and Frank Hutter},
booktitle={International Conference on Learning Representations},
year={2019},
url={https://openreview.net/forum?id=Bkg6RiCqY7},
}

@inproceedings{kharitonov-etal-2022-text,
    title = "Text-Free Prosody-Aware Generative Spoken Language Modeling",
    author = "Kharitonov, Eugene  and
      Lee, Ann  and
      Polyak, Adam  and
      Adi, Yossi  and
      Copet, Jade  and
      Lakhotia, Kushal  and
      Nguyen, Tu Anh  and
      Riviere, Morgane  and
      Mohamed, Abdelrahman  and
      Dupoux, Emmanuel  and
      Hsu, Wei-Ning",
    editor = "Muresan, Smaranda  and
      Nakov, Preslav  and
      Villavicencio, Aline",
    booktitle = "Proceedings of the 60th Annual Meeting of the Association for Computational Linguistics (Volume 1: Long Papers)",
    month = may,
    year = "2022",
    address = "Dublin, Ireland",
    publisher = "Association for Computational Linguistics",
    url = "https://aclanthology.org/2022.acl-long.593/",
    doi = "10.18653/v1/2022.acl-long.593",
    pages = "8666--8681"
}

@misc{radford2022whisper,
  doi = {10.48550/ARXIV.2212.04356},
  url = {https://arxiv.org/abs/2212.04356},
  author = {Radford, Alec and Kim, Jong Wook and Xu, Tao and Brockman, Greg and McLeavey, Christine and Sutskever, Ilya},
  title = {Robust Speech Recognition via Large-Scale Weak Supervision},
  publisher = {arXiv},
  year = {2022},
  copyright = {arXiv.org perpetual, non-exclusive license}
}

@inproceedings{saeki22c_interspeech,
  title     = {UTMOS: UTokyo-SaruLab System for VoiceMOS Challenge 2022},
  author    = {Takaaki Saeki and Detai Xin and Wataru Nakata and Tomoki Koriyama and Shinnosuke Takamichi and Hiroshi Saruwatari},
  year      = {2022},
  booktitle = {Interspeech 2022},
  pages     = {4521--4525},
  doi       = {10.21437/Interspeech.2022-439},
  issn      = {2958-1796},
}

@article{DBLP:journals/corr/SuniAV15,
  author       = {Antti Suni and
                  Daniel Aalto and
                  Martti Vainio},
  title        = {Hierarchical Representation of Prosody for Statistical Speech Synthesis},
  journal      = {CoRR},
  volume       = {abs/1510.01949},
  year         = {2015},
  url          = {http://arxiv.org/abs/1510.01949},
  eprinttype    = {arXiv},
  eprint       = {1510.01949},
  bibsource    = {dblp computer science bibliography, https://dblp.org}
}

@InProceedings{Rombach_2022_CVPR,
    author    = {Rombach, Robin and Blattmann, Andreas and Lorenz, Dominik and Esser, Patrick and Ommer, Bj\"orn},
    title     = {High-Resolution Image Synthesis With Latent Diffusion Models},
    booktitle = {CVPR},
    month     = {June},
    year      = {2022},
    pages     = {10684-10695}
}

@inproceedings{klimkov19_interspeech,
  title     = {Fine-Grained Robust Prosody Transfer for Single-Speaker Neural Text-To-Speech},
  author    = {Viacheslav Klimkov and Srikanth Ronanki and Jonas Rohnke and Thomas Drugman},
  year      = {2019},
  booktitle = {Interspeech 2019},
  pages     = {4440--4444},
  doi       = {10.21437/Interspeech.2019-2571},
  issn      = {2958-1796},
}
}
\end{document}